\documentclass[11pt]{article}
\usepackage{geometry}              
\usepackage{graphicx}
\usepackage{amssymb}
\usepackage{epstopdf}
\usepackage{amsmath}
\usepackage{stmaryrd}
\usepackage{authblk}
\usepackage{hyperref}
\usepackage[numbers, sort, compress, comma, square]{natbib}
\usepackage{booktabs}
\usepackage{array}
\newcommand{\pd}[2] {\frac{\partial #1}{\partial #2}}
\newcommand{\od}[2] {\frac{d #1}{d #2}}

\title{Buoyancy and Marangoni Effects on Horizontal Ribbon Growth}
\author[1]{Nojan Bagheri-Sadeghi}
\author[1]{ Brian T. Helenbrook\footnote{bhelenbr@clarkson.edu}}
\affil[1]{Department of Mechanical \& Aerospace Engineering, Clarkson University, Potsdam, NY 13699-5725, United States}
\date{\today}                                           

\begin{document}
\maketitle
\begin{abstract}
Unsteady simulations of horizontal ribbon growth of silicon were performed that included both Marangoni and buoyancy effects. A chaotic flow was observed dominated by strong Marangoni-driven jets emerging near the local temperature minima on the free surface.  This oscillatory flow caused the vertical position of the leading edge of the sheet to fluctuate, resulting in corrugations on the top surface of the ribbon. Additionally,   larger amplitude and wavelength nonuniformities appeared on the bottom of the sheet resulting in a sheet with varying thickness. Lastly, the unsteady flow caused temporal variations in growth rate, which when converted to distance using the pull speed, matched the wavelengths observed on the top surface. All three of these phenomena have been observed experimentally: The median of the surface wavelengths and amplitudes decreased with increasing temperature sensitivity of surface tension and had wavelengths on the same order as  experiments for a sensitivity corresponding to uncontaminated silicon. Oscillations in growth rate have been observed using passive antimony demarcation and thickness variations have been measured after sheet removal. These results indicate that the chaotic flow makes producing thin uniform sheets using HRG challenging.

\end{abstract}

\section{Introduction}

Horizontal ribbon growth (HRG) has been studied for several decades with the aim of producing lower-cost silicon sheets for solar cells than the Czochralski method, which involves losses due to squaring and sawing the ingots~\cite{Bleil:1969, Zoutendyk:1978, Zoutendyk:1980, Kudo:1980a, Glicksman:1983, Bates:1984, Thomas:1987, Daggolu:2012, Daggolu:2013, Daggolu:2014,  Helenbrook:2015, Helenbrook:2016, Kellerman:2016, Ke:2017,Sun:2018, Sun:2020, Pirnia:2021, Pirnia:2022, Daggolu:2022}. A major issue in the successful growth of silicon sheets by HRG is achieving steady conditions so that a sheet of constant thickness can be produced. 

Flow instabilities can pose a major challenge in achieving such steady conditions in crystal growth from melts. The importance of  flow instabilities due to buoyancy and surface tension gradients (Marangoni effects) was first investigated in floating zone (FZ) crystal growth~\cite{Chang:1975, Chang:1976,Schwabe:1978}. Such flow instabilities, which can cause striations on the grown crystal, was investigated numerically by Chang and Wilcox~\cite{Chang:1975, Chang:1976} and demonstrated experimentally by Schwabe et al.~\cite{Schwabe:1978}. Schwabe et al.~\cite{Schwabe:1978} studied buoyancy and Marangoni convection due to both temperature gradients (thermocapillary effect) and concentration  gradients (solutocapillary effect) and showed that oscillatory buoyancy-Marangoni convection can dominate the flow. Furthermore, Schwabe et al.~\cite{Schwabe:1979} and Chun and Wuest~\cite{Chun:1979} performed experiments on  FZ with small Bond numbers (ratio of gravity to surface tension forces) and showed the existence of steady Marangoni convection up to a critical Marangoni number (ratio of Marangoni convection to thermal diffusion) beyond which the flow became unsteady. A review of Marangoni effects in various crystal growth methods can be found in Ref.~\cite{Tsukada:2015}.  

Schwabe et al.~\cite{Schwabe:1978, Schwabe:1979} and Bates and Jewett~\cite{Bates:1984} noted that flow instabilities due to buoyancy and surface tension gradients could lead to variations in heat flux during HRG. Daggolu et al.~\cite{Daggolu:2012, Daggolu:2014}  developed a numerical model of HRG including buoyancy, Marangoni, and free surface motion but neglecting the kinetics of solidification. They reported strong Marangoni flows and weaker buoyancy-driven ones, but still steady solutions. 

Helenbrook et al.~\cite{Helenbrook:2016} developed a model of  HRG that included Marangoni effects and the kinetics of solidification but neglected  buoyancy effects. They showed that the inclusion of solidification kinetics is essential to accurately predict the faceted solidification near the triple junction point (TJP), where significant supercooling was observed. They also found steady solutions with flow speeds induced by Marangoni stresses two orders of magnitude larger than the pull speeds. 

In experiments by Kellerman et al.~\cite{Kellerman:2016}, corrugations on the top surface of the ribbon were observed  with a wavelength of roughly $10\ \mathrm{\mu m}$ for the same  setup modeled by Helenbrook et al.~\cite{Helenbrook:2016}. They attributed these ridges on the surface to solidification kinetics (i.e. alternating slow facet growth and fast roughened growth) through a heuristic limit cycle theory. They ruled out flow instabilities due to Marangoni and buoyancy forces because they postulated that the corrugation wavelengths caused by flow instabilities should vary in proportion to the pull speed and this was not observed.  

Sun et al.~\cite{Sun:2018}, duplicated the model of Helenbrook et al.~\cite{Helenbrook:2016} in COMSOL$^{\textcircled{\scriptsize R}}$ and observed a chain of vortices in their steady solutions due to Marangoni effect, similar to that reported by Helenbrook et al.~\cite{Helenbrook:2016}. They noted that these vortices became stronger as the cooling heat flux increased. Sun et al.~\cite{Sun:2020}, then simulated HRG in a simplified model with no solidification kinetics  or  realistic solid-liquid interface, but included buoyancy in addition to Marangoni effects and looked into the unsteady solution and oscillations caused by flow instabilities. Their results indicated  Marangoni and buoyancy can cause oscillations in velocity and temperature with little dependence on the pull speed.

The main purpose of this paper is to investigate the flow during HRG due to the combination of buoyancy and Marangoni effects, and see if we can explain some of the experimental observations~\cite{Helenbrook:2016, Kellerman:2016, Daggolu:2022}. Most previous models did not include kinetics which changes the temperature field significantly~\cite{Zoutendyk:1978, Zoutendyk:1980, Kudo:1980a, Glicksman:1983, Thomas:1987, Daggolu:2012, Daggolu:2013, Daggolu:2014, Ke:2017}.   Our own previous work did not include buoyancy~\cite{Helenbrook:2016}. We also note that the surface tension temperature sensitivity coefficient used in our own previous work and others~\cite{Sun:2018} was probably too low as the measured value is highly sensitive to the presence of oxygen and other impurities~\cite{Eustathopoulos:2013}.  A numerical model of the experiments reported by Kellerman et al.~\cite{Kellerman:2016}, similar to the work of Helenbrook et al.~\cite{Helenbrook:2016}, was employed with buoyancy and TJP growth angle physics added. The results, most of which are compared to experimental observations~\cite{Helenbrook:2016, Kellerman:2016, Daggolu:2022}, include the fluid dynamics, surface corrugations, growth rate variations, and changes in thickness

\section{Methods}
\subsection{Solidification Model}
The numerical model was set up similar to~\cite{Helenbrook:2016}, with a few changes discussed below, to simulate the experimental results reported in~\cite{Helenbrook:2016, Kellerman:2016,Daggolu:2022}. The experimental setup of Refs.~\cite{Helenbrook:2016, Kellerman:2016} is composed of replenishment, growth, thickness control, separation, and removal of parts consecutively. Here, only the growth region of the experimental setup was modeled. 

\begin{figure}[tb]
		\begin{center}
      			\includegraphics[width=0.7\linewidth]{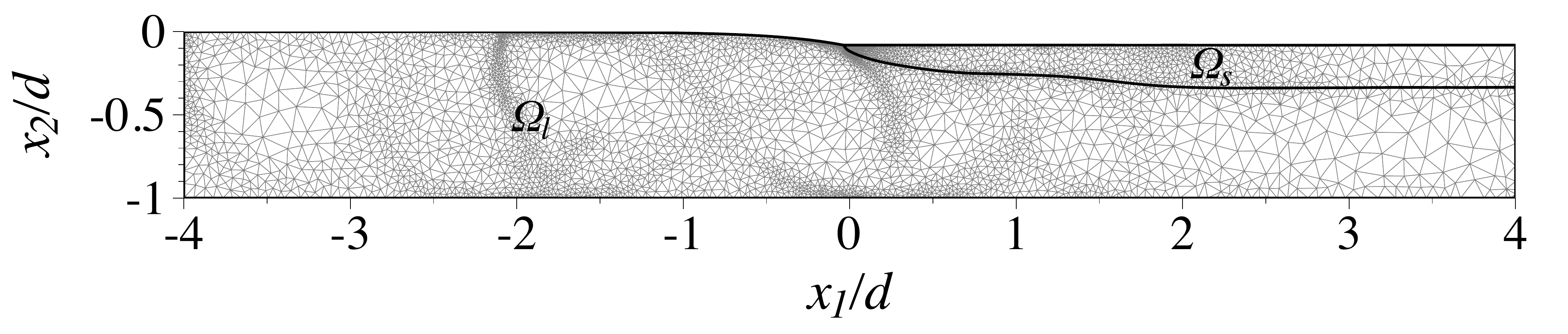}
  		\end{center}
   		\caption{Domain and an adapted mesh for the case with a pull speed of 0.5 mm/s. The actual mesh resolution is four times finer because of the quartic basis functions used on each triangular element. }
   		\label{fig:domainNMesh}
\end{figure}

A schematic of the growth region and an adapted mesh composed of a liquid region, $\Omega_l$, and a solid silicon region,  $\Omega_s$ is shown in Fig.~\ref{fig:domainNMesh}. The melt depth, $d$, in the experiments and in all of the following results was $13\  \mathrm{mm}$. At the center of the domain a cold helium slot jet impinges on the molten silicon to maintain the growth process. This is not shown in the figure but was included in the model using the heat removal boundary condition on the top surface. The domain extended $4d$ upstream and downstream of the axial position of the center of the slot jet. In the experiments, there was also a heater under the molten silicon~\cite{Helenbrook:2016, Kellerman:2013} which was included as a boundary condition in the numerical model as well. (See~\ref{sec:BC} for more details on boundary conditions).

\subsection{Governing Equations}
 The temperature field in the solid is governed by the convection-diffusion equation (written in indicial notation)
\begin{equation}
 \pd{\rho_s c_s T}{t} + \pd{\rho_s c_s T u_j }{x_j} -\pd{}{x_j} \left( k_s \pd{T}{x_j} \right) =0 
\end{equation}
where $T$ is temperature, $t$ is time, $x_j$ and $u_j$ with $j \in {1,2}$ denote the horizontal and vertical coordinates and components of velocity respectively. For all of the following, the vertical velocity in the solid, $u_{2}$, was zero while the horizontal velocity is the solid pull speed. The density, specific heat and thermal conductivity of the solid were taken as $\rho_s = 2530\ \mathrm{kg/m^3}$, $c_s = 1000\ \mathrm{J/(kg \cdot K)}$ and $k_s = 22\ \mathrm{W/(m^2 \cdot K)}$ respectively~\cite{Mito:2005}. 

The convection-diffusion equation governs the liquid part of the domain as well with subscript $s$ replaced by $l$ to show the liquid properties. For the liquid, we assumed, $c_l = c_s$, $k_l = 64\ \mathrm{W/(m^2 \cdot K)}$~\cite{Mito:2005} and the density varies linearly with temperature as
\begin{equation}
\rho_l = \rho_m + \od{\rho_l}{T} \left(T-T_m \right) 
\end{equation}
where $\rho_m = \rho_s$, $\od{\rho_l}{T} = -0.23\ \mathrm{kg/(m^3 \cdot K)}$, and $T_m = 1685\ \mathrm{K}$~\cite{Yaws:1999}. This assumes that the solid and liquid densities are equal at the equilibrium melting temperature, which simplifies the implementation of the solidification jump conditions discussed below.

 The liquid velocity components are determined from the differential forms of the conservation of mass and linear momentum of a Newtonian fluid:
\begin{equation}
\pd{\rho_l}{t} +\pd{\rho_l u_{j}}{x_j} =0 
\end{equation}
\begin{equation}
 \pd{\rho_l u_{i}}{t} +\pd{\rho_l u_{i} u_{j}}{x_j} = -\pd{p}{x_i} +\pd{\tau_{ij}}{x_j} +\rho_l g_i 
\end{equation}
where $p$ is fluid pressure, the viscous stresses are given by $\tau_{ij} = \mu \left( \pd{u_j}{x_i} +\pd{u_i}{x_j} \right)$ with the dynamic viscosity of liquid silicon $\mu = 7 \times 10^{-4}\ \mathrm{kg/(m \cdot s)}$~\cite{Mito:2005}, and $g_i$ with $i \in{1,2} $  are the gravitational acceleration components  ( $g_1 = 0$ and $g_2 = -9.8\ \mathrm{m/s^2}$).

\subsection{Solid-Liquid Interface Model}
At the solid-liquid interface, $\Gamma_I$,  conservation of mass requires:
\begin{equation}
\left \llbracket \rho (u_j - \dot{x}_j) n_j \right \rrbracket_{\Gamma_I}= 0 
\end{equation}
 where $ [\![   ]\!]$ denotes the jump across the interface, $\rho = \rho_s$ on the solid side and $\rho= \rho_l$ on the liquid side, $\dot{x}_j$ with $j \in {1,2}$ are the interface velocity components and $n_j$  are components of the outward normal pointing in opposite directions for solid and liquid. Although at the interface liquid density varies because of kinetic supercooling, it was assumed that at the interface $\rho_l=\rho_s$ and therefore liquid and solid velocities were equal. Hence, at the interface, a Dirichlet boundary condition for the velocity components of the liquid was imposed where $u_1$ was set to the pull speed and $u_2$ was set to 0.
 
Conservation of energy at the interface  states that the jump in the energy flux should be equal to the flux of energy absorbed through phase change 
\begin{equation}
\label{eq:IntEnergy}
\left \llbracket \left(\rho c T (u_j - \dot{x}_j) - k \pd{T}{x_j} \right) n_j \right \rrbracket_{\Gamma_I} = \rho_s (u_j - \dot{x}_j) n_{s,j} L_f 
\end{equation}
Where $n_{s,j}$ is the outward normal to the solid at the interface and the latent heat of fusion, $L_f$, was taken as  $ 1.8 \times 10^6\ \mathrm{J/kg}$.

The solidification kinetics at the interface was based on the model used in Ref.~\cite{Weinstein:2004} where the interface supercooled temperature is determined as
\begin{equation}
\label{eq:kinetics}
\Delta T = K(\Delta T, \theta_m) (u_j - \dot{x}_j) n_{s,j}   
\end{equation}
where $\Delta T = T - T_m$ is the temperature difference of the interface from the equilibrium melting temperature and $K(\Delta T, \theta_m)$ is the kinetic coefficient that is a function of $\Delta T$,  and the misalignment angle, $\theta_m$, from the \{111\} facet direction. It was assumed that the growth was initiated with the $[\bar{1}00]$ direction pointing upward and the $[011]$ direction aligned with the direction of growth. In this case the \{111\} plane is about $55^\circ$ from the horizontal axis. The kinetic coefficient was defined as: 
\begin{equation}
\begin{cases} 
      K= K_{2DN}, &  \sin(\theta_m) = 0, \\
      K = \left( K_{rough}^4 +K_{step}^4\right)^{1/4}, &   \sin(\theta_m) > 0,
   \end{cases}
\end{equation}
where 
\[ K_{2DN} = B^{-1} e^{\frac{-A}{|\Delta T|}}\] 
\[ K_{step} = \frac{K_{SN}}{\left | \sin(\theta)\right | +\epsilon_{step}}\]
and $A = 140\ \mathrm{K}$ and $B =1.5\times 10^{10}\ \mathrm{K\  s/m} $, $K_{rough} = 79.4\  \mathrm{K\ s/m} $, and $K_{SN} = 144\  \mathrm{K\ s/m} $~\cite{Weinstein:2004}. The value of $\epsilon_{step}$ was set to machine precision (i.e. about $10^{-16}$) to avoid division by zero.  $K_{2DN}$ models the two-dimensional nucleation mechanism of crystal growth, which we assume occurs at the TJP where the temperature is lowest as shown in~\cite{Helenbrook:2016}. The growth along a facet is dominated by step nucleation mechanism, $K_{step}$. As the misalignment angle $\theta_m$ increases, the crystal growth becomes rough on the atomic scale and the kinetic coefficient value is dominated by $K_{rough}$. The value of $K_{SN}$ was set about 90 times greater than the value in Ref.~\cite{Weinstein:2004} to avoid high sensitivity to misalignment angle that led to convergence issues. In previous work, we found that if the value of $K_{SN}$ from Ref.~\cite{Weinstein:2004} were used, the facet is slightly flatter and there is a sharper transition to roughened growth. Equation~\ref{eq:IntEnergy} coupled with the solidification kinetics  was used to determine the normal interface velocity. 

\subsection{Boundary Conditions}
\label{sec:BC}

At the left side of the domain, the inlet velocity components and temperature were specified as an isothermal channel flow $u_1 = u_{s,1} \left( 1 - \left( \frac{x_2}{d} \right)^2 \right)$, $u_2 = 0$ and $T - T_m =5\ \mathrm{K}$. At the right of the domain, an outflow condition  was imposed for the liquid by setting a zero total stress. A condition of zero heat flux was applied for both solid and liquid on the right side of the domain. 

A growth angle of $\theta_g = 11^\circ$ was imposed by constraining the direction of  motion of  the TJP relative to the normal to the free surface, such that 
\begin{equation}
\frac{\left( \dot{x}_{TJP,j} -u_{s,j} \right)  n_j}{\sqrt{\left( \dot{x}_{TJP,j} -u_{s,j} \right) \left( \dot{x}_{TJP,j} -u_{s,j} \right)  }} = \sin \theta_g
\end{equation}
 where $\left( \dot{x}_{TJP,j} -u_{s,j} \right) $  are components of mesh velocity at the triple junction point relative to the solid motion~\cite{Surek:1976, Eustathopoulos:2010} .  If $\theta_g = 0$, this forces the solidification at the TJP to grow tangent to the free surface. A growth angle of $11^\circ$ results in the free surface approaching the TJP from an $11^\circ$ incline relative to horizontal in the steady state case.
 
 As the height of the TJP varied, this varying height was translated with the pull speed along the top surface of the solid. Therefore, corrugations could be observed along the top surface of the solid. Because the mesh became coarser away from the TJP, the smaller wavelengths became unresolved on the top surface of the solid. To fix this issue, the corrugations on the top surface of the solid were reconstructed analytically from the variations in the position of the TJP. 

The flow boundary conditions at the free surface of the liquid were the kinematic condition that there is no flow through to the interface
 \begin{equation}
\left(u_j - \dot{x}_j \right)  n_j = 0 
 \end{equation}
 and the stress on the free surface was defined to be equal to stresses due to the surface curvature and the temperature dependence of surface tension (i.e. the Marangoni effect):
 \begin{equation}
     -p n_i +\tau_{ij} n_j = \pd{\sigma(T) t_i}{s}  
 \end{equation}

  where $t_i$ denote the components of the unit tangent vector to the free surface and the surface tension, $\sigma$,  is a function of temperature.  $\sigma$ was taken as  $\sigma = \sigma_0 +\od{\sigma}{T} (T - T_m)$,  $\sigma_0$ had a value of  $0.735\ \mathrm{N/m}$~\cite{Mito:2005} and two values of the surface tension temperature sensitivity were studied. $\od{\sigma}{T} = -4\times10^{-4}$ which corresponds to pure silicon in argon atmosphere~\cite{Kobatake:2015} and a reduced value of $\od{\sigma}{T} = -1\times10^{-4}$ which corresponds to presence of some impurities in the melt~\cite{Eustathopoulos:2013}.

The thermal boundary condition on the top of the domain, for solid and liquid, was a specified heat flux as
\begin{equation}
q =q_c +q_r 
\end{equation}
where the convective heat flux of helium, $q_c$ was modeled as
\begin{equation}
q_c = q_{base} + q_{peak} \left((1- \zeta) 2^{-(x/w)^2}  +\zeta 2^{-(x/w_b)^2} \right)
\end{equation}
where $q_{base}$, $q_{peak}$, $\zeta$, $w$ and $w_b$ are curve fit coefficients. The curve fit was based on results of three  ANSYS Fluent$^{\textcircled{\scriptsize R}}$ 16.2 simulations of the  slot jet for different helium flow rates~\cite{Helenbrook:2016}. For all the cases here, $q_{base} = 164\ \mathrm{kW/m^2}$ represents the conductive heat transfer between the melt and helium, $\zeta = 0.55$, $w_b = 1.44\ \mathrm{mm}$ and values of $q_{peak}$, and $w$ are given in Table~\ref{tbl:cases} along with pull speeds and values of $\od{\sigma}{T}$ of these cases. Note that the heat fluxes for cases 1, 2 and 4 were based on the experimental work of Kellerman et al. ~\cite{Kellerman:2016} with the helium flow rate of $Q_{He} = 1.9\ \mathrm{L/min}$ and $Q_{He} = 2.5\ \mathrm{L/min}$ respectively. Case 3 had the same $q_c$ as the first case of Helenbrook et al.~\cite{Helenbrook:2016} with $Q_{He} = 5.0\ \mathrm{L/min}$.

\begin{table}[tb]
\centering
\caption{Pull speeds, temperature sensitivities of surface tension and curve fit parameters of the helium jet heat flux,  $q_c$, for cases studied \label{tbl:cases}}
\begin{tabular}{ cccccc}
\toprule
Case & \textbf{$u_{s,1}\ \mathrm{\left(\frac{mm}{s}\right)}$} &    $\od{\sigma}{T}\ \mathrm{ \left(\frac{N}{m \cdot K} \right)}$ & \textbf{$q_{peak}\ \mathrm{\left(\frac{MW}{m^2}\right)}$}  & \textbf{$w\ \mathrm{(mm)}$}  \\
\midrule
	 1	&  0.5 	& $ 1\times 10^{-4}$	&  0.95 	&	 0.68   \\
	 2 	&   0.7   	& $ 1\times 10^{-4}$	&  1.26 	&	 0.59    \\
	3 	& 1  		& $ 1\times 10^{-4}$	&  2.53	&  0.42  \\
	 4 	&   0.7   	& $ 4\times 10^{-4}$	&  1.26 	&	 0.59    \\
\bottomrule
\end{tabular}

\end{table}

\begin{figure}[tb]
		\begin{center}
      			\includegraphics[width=0.4\linewidth]{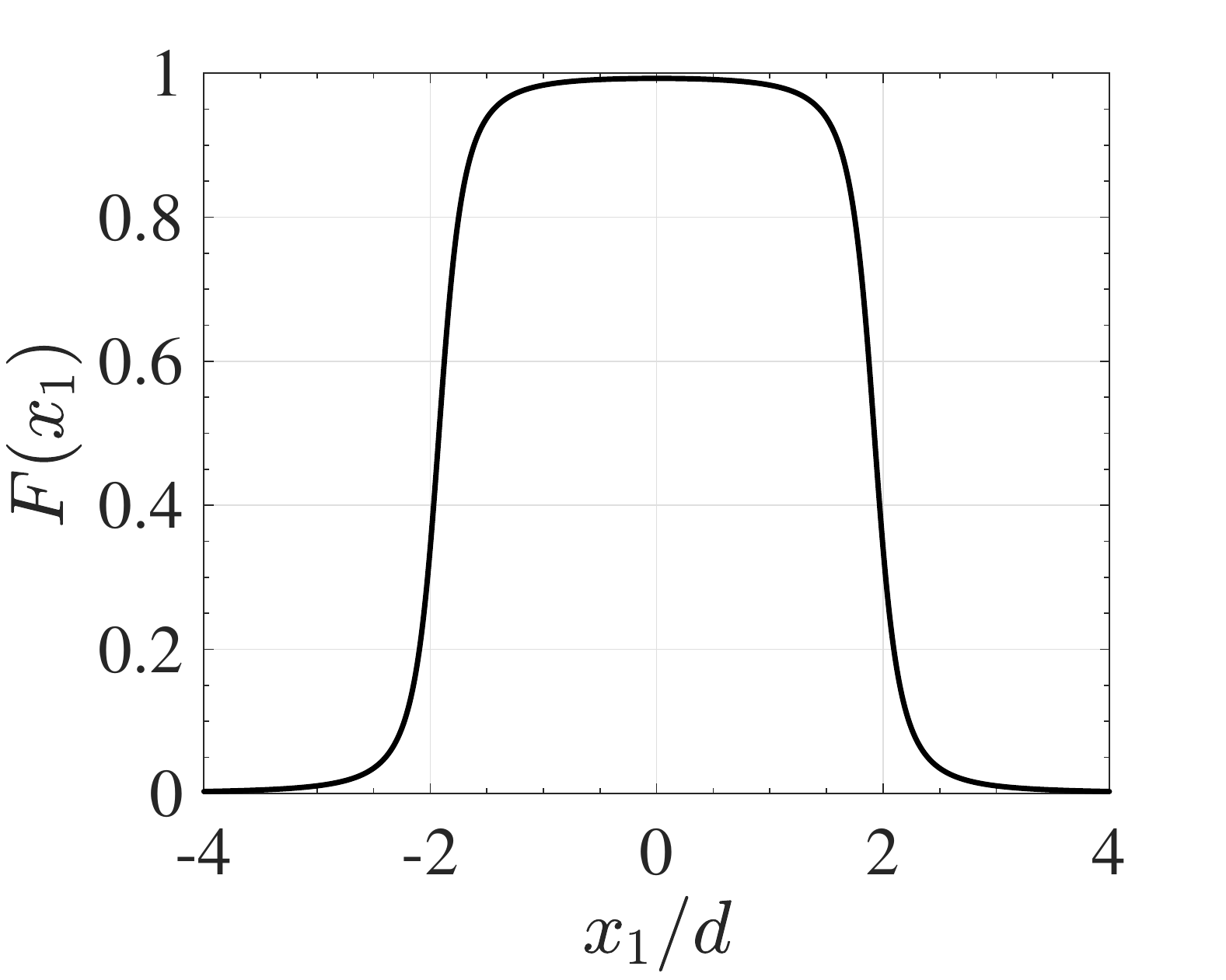}
  		\end{center}
   		\caption{The view factor function $F(x_1)$}
   		\label{fig:VF}
\end{figure}

The radiation heat flux $q_r$ between the silicon and the water cooled block that contained the helium slot jet was modeled assuming the block to be a horizontal surface centered above the domain. The effect of the growth angle on surface shape was neglected (i.e. the liquid and solid surfaces were assumed to be flat at $x_2 = 0$). The radiative heat flux was calculated as
\begin{equation}
q_r = \epsilon \sigma_b F(x_1) (T_m^4 - T_c^4) 
\end{equation}
where $\epsilon$ is the emissivity and has different values of $\epsilon_l = 0.2$ and $\epsilon_s = 0.6$ for liquid and solid  respectively. The Stefan-Botlzman constant is denoted as $\sigma_b$ and $F(x_1)$ is the view factor between the water-cooled block at $T_c = 300\ \mathrm{K}$ and the top surface defined as~\cite{Howell:2016}
\begin{equation}
 F(x_1) = \frac{\sin \phi_2 - \sin \phi_1}{2} 
\end{equation}
where 
\[ \sin \phi_1 = \frac{-w_r/2 - x_1 }{\sqrt{\left( -w_r/2 - x_1 \right)^2 +h_{r}^2}} \]
\[ \sin \phi_2 = \frac{w_r/2 - x_1 }{\sqrt{\left( w_r/2 - x_1 \right)^2 +h_{r}^2}} \]
where the width of the block was $w_r = 5\ \mathrm{cm}$ and the height of the block from the top of the melt (i.e. from $x_2 = 0$)  was $h_r = 3\ \mathrm{mm}$. The behavior of $F(x_1)$ is shown in Fig.~\ref{fig:VF}.

At the bottom of the domain, a no-slip boundary condition and specified heat flux were imposed.   The stabilizing heat flux from the bottom was set to match case 1 from Ref.~\cite{Helenbrook:2016}. In the experiment, a heater was located under the melt with about the same width as the water cooled block. To model this, the bottom heat flux was given in $\mathrm{kW/m^2}$ as 
\begin{equation}
q_b = 244.4  F(x_1) 
\end{equation}
where the view factor function $F(x_1)$ was used as a convenient function for confining the heat addition to the region below the water cooled block.

\subsection{Numerical Method}
A third-order accurate, 4-stage, L-stable diagonally implicit Runge-Kutta (DIRK) scheme was used for time advancement. A high order finite element method (hp-FEM) using fourth-degree basis functions on triangular elements was used to obtain the numerical solution in space~\cite{Helenbrook:2018}. The hp-FEM used the streamline-upwind Petrov-Galerkin (SUPG) stabilization approach for the equal order pressure and velocity approximation space~\cite{Helenbrook:2018}. An arbitrary-Lagrangian-Eulerian (ALE) moving mesh method was used to track the solid-liquid interface. the liquid free surface and the solid free surface while adapting the mesh to maintain quality and accuracy as detailed in Ref.~\cite{Helenbrook:2018}. Mesh adaptation was based on achieving a uniform target truncation error over the domain. We also put a restriction on the minimum resolution, $l_{min}$, to avoid exessive refinement near singular points. A transient mesh with mesh adaptation is shown in Fig.~\ref{fig:domainNMesh}. 

Initial conditions were chosen as detailed in appendix~\ref{sec:IC}. A steady solution without Marangoni and buoyancy effects was first obtained during the process (discussed in appendix~\ref{sec:SM}). For cases 1 to 3 the results were then obtained at a constant time step of $\Delta t = \frac{l_{min}}{u_{s,1}}$ where $ l_{min} = 5\ \mathrm{\mu m}$. The time-stepping was done for a total time of $\frac{8d}{u_{s,1}}$.  The time step was set so that the corrugations on the top surface of the solid travel about $5\ \mathrm{\mu m}$ at each time step, allowing observation of wavelengths as small as $10-15\ \mathrm{\mu m}$, which were reported by Kellerman et al.~\cite{Kellerman:2016}. 

For case 4, the $l_{min}$ and $\Delta t$ were reduced by a factor of 4. The simulation for this case was continued from the last time step of case 2 and  $\od{\sigma}{T}$ was increased to $4\times 10^{-4} \  \mathrm{N/(m \cdot K)}$. At the increased $\od{\sigma}{T}$  the maximum velocity in the flow achieved on the free surface increased by a factor of about two. This was achieved after only 10 time steps indicating the effect of new value of $\od{\sigma}{T}$ has been established on the free surface. The results presented for case 4 ignored the first 300 time steps to discard data affected by transition in $\od{\sigma}{T}$. Convergence at $\od{\sigma}{T} = 4\times 10^{-4} \  \mathrm{N/(m \cdot K)}$ proved to be more difficult and the time-stepping was only continued for about $\frac{1.4d}{u_{s,1}}$. Because the wavelengths were smaller for this case, more waves were detected than in case 2 and therefore the data was more statistically converged in terms of the median of wavelengths and amplitudes.

\section{Results and Discussion}
\label{sec:results}
\subsection{Flow Dynamics}
\label{sec:flow}

The flow field was unsteady and did not approach a steady solution. Although the maximum velocity magnitude due to the thermocapillary effect was an order of magnitude greater than the maximum velocity due to buoyancy, the inclusion of buoyancy in the model was essential to observe the unsteadiness. With buoyancy in the model, even with no Marangoni stresses, the flow was unsteady at all pull speeds. Helenbrook et al.~\cite{Helenbrook:2016} reported steady laminar solutions of a similar model with $\od{ \sigma}{ T} = -7 \times 10^{-5}\ \mathrm{N/(m \cdot K)}$  from simulations when buoyancy effects were neglected. 

\begin{figure}[tbp]
		\begin{center}
      			\includegraphics[width=0.85\linewidth]{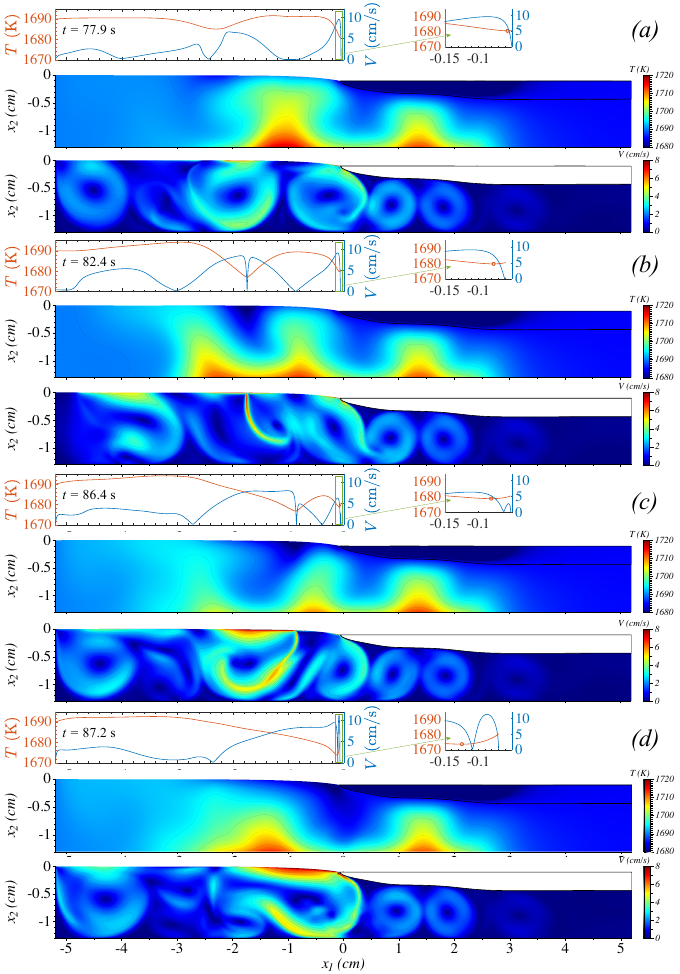}
  		\end{center}
   		\caption{Four consecutive snapshots of profiles of temperature and velocity along the   melt surface and corresponding  temperature and velocity contours  for case 1 ($\od{\sigma}{T} = 1 \times 10^{-4}\ \mathrm{N/(m \cdot K)}$ and $u_{s,1} = 0.5\ \mathrm{mm/s}$). The contours of temperature and velocity are respectively 5 K and 0.5 cm/s apart. The marker in the zoomed-in views identifies the point of minimum temperature. See \href{https://youtu.be/_kwhDn96Dc8}{Video~1}.  }
   		\label{fig:flow}
\end{figure}

Fig.~\ref{fig:flow} shows four consecutive snapshots of the unsteady temperature and velocity fields. The line plots show the velocity magnitude and temperature on the free surface aligned with the subsequent contour plots. \href{https://youtu.be/_kwhDn96Dc8}{Video 1} shows a movie of the flow in a similar manner to Fig.~\ref{fig:flow}. In our unsteady simulations,  a supercooled region was always present in front of the TJP and there was a point of minimum temperature on the surface in this region near the TJP.  This point is identified by a circular marker in the zoomed-in views of the line plots shown to the right at the full line plots in Fig.~\ref{fig:flow}. At this point,  surface tension attained its maximum value and pulled the melt at the surface from both sides.

This pull often created a small counterclockwise vortex, between this point and the TJP,  similar to what was reported in steady solutions of Helenbrook et al.~\cite{Helenbrook:2016}  (see the zoomed-in views of Fig. 10 in~\cite{Helenbrook:2016} or the zoomed-in view of velocity magnitude contour plot in \href{https://youtu.be/_kwhDn96Dc8}{Video 1} at time $t = 86.7 \mathrm{s}$). The small vortex quickly rolled up into a jet and merged with the large clockwise vortex beneath the TJP. This vortex circulated cold fluid downward and warm fluid upward creating the alternating cold and hot temperature fields seen in Fig.~\ref{fig:flow}a-d.

Generally, the minimum supercooled surface temperature fluctuated and as it became colder or warmer, it moved further upstream or downstream respectively and the TJP followed it. It is notable that the point of the high-velocity jet emerging near the TJP in zoomed-in surface profiles of Fig.~\ref{fig:flow}  follows the point of minimum temperature with a lag. This time lag between the point of maximum surface tension and jet position keeps disturbing the velocity field that in turn disturbs the temperature field as it changes. Such interactions between temperature and velocity fields can contribute to the chaotic flow field and high-frequency changes in the TJP position. 

Downstream to the right of the large clockwise vortex beneath the TJP, there were three other large vortices rotating in counterclockwise, clockwise, and clockwise directions respectively. Ordinarily, three other large vortices could be discerned upstream of the large vortex beneath the TJP that from the most upstream one were rotating in counterclockwise, clockwise, and counterclockwise directions respectively. Buoyancy fed energy into these large vortices as it pulled the colder melt from the surface or just beneath the sheet downward and pushed the hotter melt near the bottom upward.

Additionally, there were one or more regions of low temperature further upstream of the TJP. Often, there was a point of minimum temperature on the surface of these regions and as the unsteady temperature field evolved, these regions could attain supercooled temperatures temporarily. At such local maxima of surface tension, the melt was pulled from both sides. In some cases, initially, a temporary small vortex formed at these points that returned the cold melt to the surface. These small vortices were short-lived and rolled up into a jet streaming into the melt. Fig.~\ref{fig:flow}b shows a clear formation of such a jet around  $x_1 = -1.8\ \mathrm{cm}$. 

The position of jet ejection into the melt in the velocity profiles along the surface can be discerned as points where velocity sharply decreases towards zero similar to a stagnation point. Notably, the position of the jet at the surface closely follows the point of minimum temperature in Fig~\ref{fig:flow}. As the position of the minimum temperature changed, the jet moved back and forth. The temperature at that point increased as the warmer melt moved towards the point of minimum temperature or as the point moved away from the middle of the domain with the maximum cooling. Conversely, movement of the point further upstream away from the point of maximum cooling by the helium jet, decreased its temperature. Eventually, jets were either pulled towards the TJP or away from it. If pulled towards the TJP they often became stronger and merged with the jet streaming at the minimum temperature near the TJP into a stronger cold jet flowing into the crucible. The movement of a jet and merging with the jet at the TJP are shown in Figs.~\ref{fig:flow}c and~\ref{fig:flow}d. If moved away from the TJP, such jets became weaker and eventually disappeared. Additionally, these cold high velocity jets streamed into the melt and disrupted the temperature field and large vortices beneath and upstream of the TJP. Such disruptions in the flow field are shown in Figs.~\ref{fig:flow}b to~\ref{fig:flow}d .

For the case 4 where $\od{\sigma}{T} = 4 \times 10^{-4}\  \mathrm{N/(m \cdot K)}$, compared to case 2 with $\od{\sigma}{T} = 1 \times 10^{-4}\  \mathrm{N/(m \cdot K)}$, the maximum velocity induced by surface tension gradients increased by a factor of about 2.3 from an average maximum velocity of about 10 cm/s to 23 cm/s. Therefore, in this case, jets of higher velocity streamed into the crucible, and reduced the time scales of flow oscillations.  

Comparing cases 1 to 3 with  $\od{\sigma}{T} = 1 \times 10^{-4}\  \mathrm{N/(m \cdot K)}$, increasing the pull speed from 0.5 to 1 mm/s, no significant change in flow characteristics was observed. This was expected as the flow field was dominated by buoyancy and Marangoni effects inducing velocity magnitudes much larger than the pull speed.

As these dynamics in the flow field caused large changes in velocity magnitude and direction near the TJP, the height and horizontal position of the TJP varied. As the leading edge of the sheet was pulled with varying heights and positions, corrugations were formed on the top surface of the solid sheet. Similarly,   the solidification interface was also affected by this dynamic flow field resulting in large variations in the shape of the sheet on the bottom and therefore the sheet thickness.
  
  \subsection{Corrugations  on the Top Surface of the Sheet}
  \label{sec:corr}
\begin{figure}[tb]
		\begin{center}
      			\includegraphics[width=\linewidth]{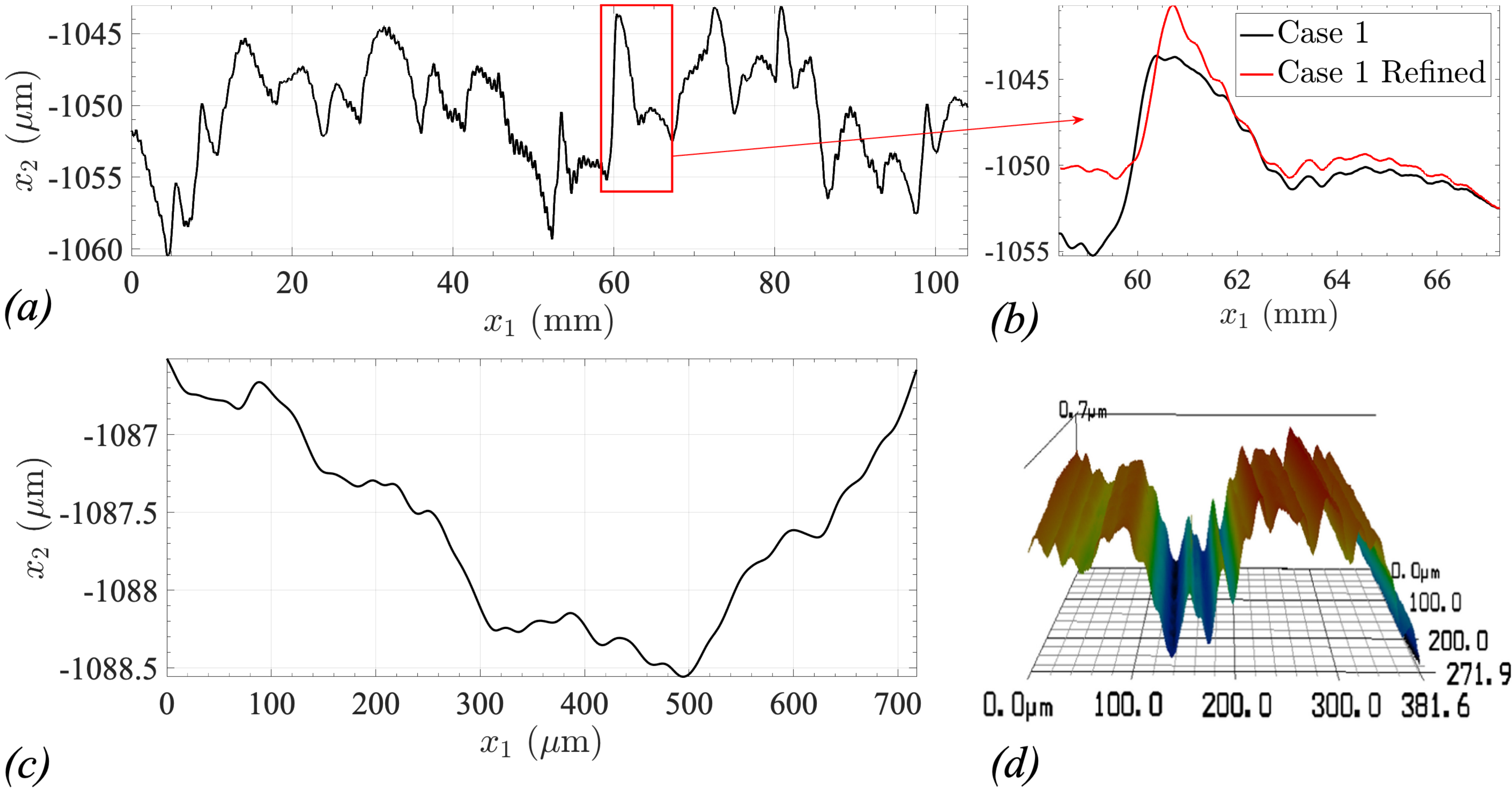}
  		\end{center}
   		\caption{Corrugations on the top surface of solid: (a) Corrugations from simulations for case 1 with $\od{\sigma}{T} = 1 \times 10^{-4}\  \mathrm{N/(m \cdot K)}$ and $u_{s,1} = 0.5\ \mathrm{mm/s}$ (b) A zoomed-in view of corrugations of case 1 and the results for case 1 refined with a finer mesh and time step halved (c) A part of surface corrugations for   case 4 with $\od{\sigma}{T} = 4 \times 10^{-4} \  \mathrm{N/(m \cdot K)}$ and $u_{s,1} = 0.7\ \mathrm{mm/s}$ (d) Experimental results of Kellerman et al.~\cite{Kellerman:2016} using confocal microscopy.    }
   		\label{fig:Corrugations}
\end{figure}

Corrugations observed for case 1 are shown in Fig.~\ref{fig:Corrugations}a. A zoomed-in view  is shown in Fig.~\ref{fig:Corrugations}b along with the results for case 1 with a finer mesh and a smaller time step to assess the sensitivity of the results to spatial and temporal resolutions. The simulation for the refined mesh was started from a solution of case 1 and was repeated for a portion of simulation time. The time step was reduced by a factor of two, the truncation error target reduced by an order of magnitude (resulting in an increase in the average number of degrees of freedom of the mesh by  a factor of almost two), and $l_{min}$ was reduced by half.   Note that the surface corrugations were pulled to the right and thus in Fig.~\ref{fig:Corrugations}b the initial point of refined simulations is at $x_1 = 67.2\ \mathrm{mm}$. As the simulation advanced in time the deviation between the resulting corrugations of original and refined cases increased. Considering the chaotic flow field dynamics discussed in~\ref{sec:flow} this is not surprising. However, Fig.~\ref{fig:Corrugations}b indicates that the average wavelengths are slightly smaller in the refined case suggesting that more refined spatial and temporal simulations would converge to results with slightly smaller wavelengths. Due to the singularity at the TJP, despite the high order spatial and temporal schemes used, the results at the TJP can converge slowly~\cite{Pirnia:2021}. 

The surface corrugations from the last portion of the simulation of case 4 with $\od{\sigma}{T} = 4\times 10^{-4}\  \mathrm{N/(m \cdot K)}$ are shown in Fig.~\ref{fig:Corrugations}c and the experimental results using confocal microscopy from Kellerman et al.~\cite{Kellerman:2016} are reproduced in Fig.~\ref{fig:Corrugations}d for comparison. Note the change in units of the $x_1$-axis  to $\mathrm{\mu m}$ in Figs.~\ref{fig:Corrugations}c and~\ref{fig:Corrugations}d from $\mathrm{mm}$ in Figs.~\ref{fig:Corrugations}a and~\ref{fig:Corrugations}b. Also,  $x_2$ varies in a range of about $2\ \mathrm{\mu m}$ and $0.7\ \mathrm{\mu m}$ in Figs.~\ref{fig:Corrugations}c  and~\ref{fig:Corrugations}d respectively.

Statistics of the wavelengths including the number of detected wavelengths $N$, median and mean of the wavelength of corrugations, $\tilde{\lambda}$ and $\overline{\lambda}$, minimum and maximum wavelengths, $\lambda_{min}$ and $\lambda_{max}$ and the median and mean of peak-to-peak wave amplitudes,  $\tilde{A}$ and $\overline{A}$, are given in Table~\ref{tbl:Waves} for both the numerical and experimental results. Unlike the reports from~\cite{Kellerman:2016}, $\tilde{\lambda}$ does show some dependence on pull speed, however, this dependence is not consistent across different ways of measuring wavelength.  For example, $\lambda_{min}$ and $\lambda_{max}$ (the maximum and minimum distance between local extreme) show little sensitivity to pull speed while $\overline{\lambda}$ increases from case 1 to 2 but barely changes from case 2 to case 3.  For case 4, in agreement with with Fig.~\ref{fig:Corrugations}c, $\tilde{\lambda}$ assumed a much smaller value of $80\ \mathrm{\mu m}$ compared to case 2, but is still larger than the experimental values shown on the last line of the Table~\ref{tbl:Waves}.  Finally, note that  $\tilde{A}$ values in Table~\ref{tbl:Waves} are on the same order as experimental results shown in Fig.~\ref{fig:Corrugations}d from Kellerman et al.~\cite{Kellerman:2016}. For cases 1 to 3, the values of $\tilde{A}$ is about three times larger than the experimental corrugations shown in Fig.~\ref{fig:Corrugations}d and for case 4, $\tilde{A}$ is 42\% greater than the experimental value.

The median wavelengths of cases 1 to 3 correspond to TJP vertical oscillations of about 1 Hz for all three cases.  As mentioned in section~\ref{sec:flow} and can be seen in Fig.~\ref{fig:flow}a, there is a large vortex beneath the TJP with a diameter of the same size as the depth of the melt.  Noting the velocity scale of about 3.5 cm/s, the turnover time of this vortex matches the observed frequency and could be the reason for the observation of increasing wavelength proportional to pull speed. In case 4,   there was a stronger jet similar to that shown in Figs~\ref{fig:flow}b-d  near the TJP disrupting the vortex. When not disrupted by cold jets streaming from the surface, the velocity of this vortex was about 6 cm/s corresponding to a frequency of about 1.5 Hz. However, for case 4 rather than a corresponding median wavelength of about 450 $\mathrm{\mu m}$, $\tilde{\lambda}$ of about 80  $\mathrm{\mu m}$ was observed. Therefore, it seems that only some of the wavelengths corresponding to vortical structures in the flow with a specific frequency scaled with pull speed.

Kellerman et al.~\cite{Kellerman:2016} gradually increased the pull speed from 0.3 mm/s to 0.8 mm/s while increasing the cooling provided by the helium jet  in their experiment from which they concluded surface wavelengths are independent of pull speed (Similarly, we increased the corresponding cooling heat flux on the top boundary condition as detailed in~\ref{sec:BC}).  However, as they increased both the pull speed and helium jet flux, they may have caused larger Marangoni stresses near the TJP with corresponding smaller time scales such that the average wavelengths did not change significantly.  Also, although not included in our model, as the pull speed increases, the segregation of solutes in the melt increases~\cite{Daggolu:2014}. This can cause Marangoni stresses due to concentration gradients. Furthermore, the thermal Marangoni stresses could be large, similar to case 4, such that  jets streaming into the flow due to Marangoni stresses disrupted the vortical structures with specific frequencies that can result in wavelengths increasing proportional to pull speed. Finally, note that there is some variance in the experimental wavelengths as shown in Table~\ref{tbl:Waves} and the wavelengths in our results showed no clear dependence on pull speed in terms of mean, maximum or minimum wavelength.

\begin{table}[tb]
		\centering
		\caption{Pull speeds; temperature sensitivities of surface tension; number,  median, mean, minimum, and maximum of wavelengths; and the median and mean of the peak-to-peak amplitude of the surface waves}
		\label{tbl:Waves}
		\begin{tabular}{@{}ccccccccccc@{}}
		\toprule
		Case &  \textbf{$u_{s,1}\newline\mathrm{\left(\frac{mm}{s}\right)}$} &    $\od{\sigma}{T}\newline\mathrm{ \left(\frac{N}{m \cdot K} \right)}$  & N  &$\tilde{\lambda}\newline \mathrm{(\mu m)}$ & $\overline{\lambda}\newline \mathrm{(\mu m)}$ &  $\lambda_{min}\newline \mathrm{(\mu m)}$ &  $\lambda_{max}\newline \mathrm{(mm)}$ & $\tilde{A}\newline \mathrm{(\mu m)}$ & $\overline{A}\newline \mathrm{(\mu m)}$ \\ \midrule
			
		1  	&  0.5	&  $ 1\times 10^{-4}$	& 176 &  518     & 588 & 40 & 3.2	& 0.43   &  0.82\\
		
		 2	&  0.7	&  $ 1\times 10^{-4}$	& 	104 & 699  &  999 &   70&  5.3  &  0.34 & 1.20\\
		
		3	&  1	& $ 1\times 10^{-4}$	& 	97 & 1013   &   1067 &   33  &  3.9	& 0.41 & 1.61\\

		 4	&  0.7	& 	$ 4\times 10^{-4}$& 	127 &  80 &  143 &   7 &  1.2	& 0.17   &  0.75\\
	
		$\mathrm{Exp.}^*$ &  0.5	&  ---	& 	 17 &  21 & 25   & 12  & 0.065  & 0.12 & 0.16\\
		\bottomrule
	\end{tabular}
	 	\begin{flushleft}
			\small $^*$  Experimental data from Fig.~\ref{fig:Corrugations}d  \\

		\end{flushleft}
	\end{table}

\subsection{Growth Rate Variations}

\begin{figure}[tb]
		\begin{center}
      			\includegraphics[width=\linewidth]{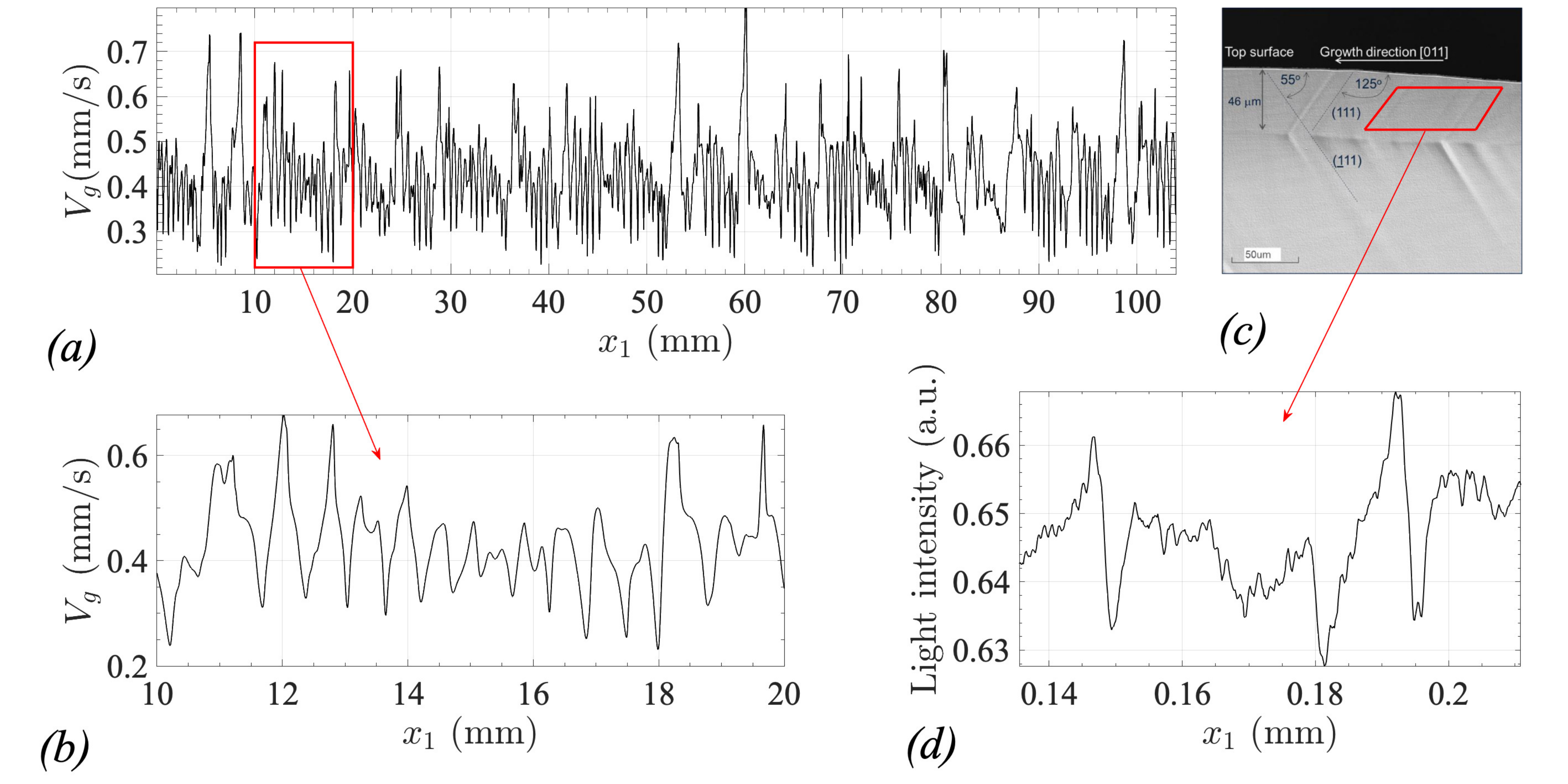}
  		\end{center}
   		\caption{Growth rate variations (a) simulation results for case 1 with $\od{\sigma}{T} = 1 \times 10^{-4} \  \mathrm{N/(m \cdot K)}$ and $u_{s,1} = 0.5\ \mathrm{mm/s}$ (b) a zoomed-in view of of case 1 (c) experimental results of Kellerman et al.~\cite{Kellerman:2016} using antimony demarcations to show regions of sharp gradients in growth rate (d)  mean light intensity normalized by maximum light intensity along the longer side of the parallelogram-shaped region }
   		\label{fig:GR}
\end{figure}

The leading edge of the ribbon is faceted. This was shown in our previous simulations~\cite{Helenbrook:2016} and also can be seen by zooming in on the TJP regions shown in Fig.~\ref{fig:flow}. The solidification. velocity of the facet can be calculated as $u_{s,1} \sin(\theta_f) + \dot{x}_{TJP,j}n_j$ where $\theta_f = 55^\circ$ is the \{111\} facet angle. The growth rate variations at the TJP for case 1 are shown in Figs.~\ref{fig:GR}a-b. The experimental results of Kellerman et al.~\cite{Kellerman:2016} obtained using a passive antimony demarcation method are shown in Fig.~\ref{fig:GR}c for comparison. The high sensitivity of antimony segregation coefficient to growth rate is used in Fig.~\ref{fig:GR}c, combined with a Wright etch~\cite{Jenkins:1977} to delineate regions of high and low antimony, as an indicator of changes in growth rate. Note that to make comparisons with experiments easier, changes of growth rate in time were mapped to their respective positions along the sheet considering the pull speed and the changing position of the TJP. 

Large gradients in light intensity in Fig.~\ref{fig:GR}c corresponds to sharp changes in growth rate. Fig.~\ref{fig:GR}d  shows the mean light intensity along the horizontal direction side of the parallelogram-shaped region in Fig.~\ref{fig:GR}c normalized by maximum light intensity. The mean light intensity was averaged along a line parallel to the smaller side of the parallelogram, which aligned with the facet. Note that the growth in the cross-section shown in Fig.~\ref{fig:GR}c was double faceted with a facet intersection point below the surface. This configuration was studied in~\cite{Pirnia:2022} but has not been included in the current model. The noisiness of the photo is reflected in the light intensity line plot. However, three regions with sharp changes in growth rate are distinguishable and they are qualitatively similar to the gradients in growth rate shown in Fig.~\ref{fig:GR}b. 

The spacings between sudden changes in growth rate experimentally observed in Fig.~\ref{fig:GR}d are similar to experimental wavelengths in Fig.~\ref{fig:Corrugations}b. Similarly, the spacings between the sharp changes in growth rate from simulations in Fig.~\ref{fig:GR}a-b are close to wavelengths obtained from our numerical model in Fig.~\ref{fig:Corrugations}a. Thus, these wavelengths scale with $\od{\sigma}{T}$ like the surface corrugations.  Furthermore, note that sudden changes in growth rate in Figs.~\ref{fig:GR}a-b can be on the same order as the steady-state growth rate itself as the growth rate sharply drops from a maximum value to a minimum value. Therefore, these large variations in growth rate can cause the experimental passive antimony demarcation observations. As both the surface corrugations and growth rate variations observed in the experiment can be explained by the chaotic flow dynamics due to Marangoni stresses and buoyancy, there seems to be no need for the heuristic limit cycle theory proposed in~\cite{Kellerman:2016, Daggolu:2022} to explain these phenomena.

\subsection{Variations in Thickness}

\begin{figure}[tb]
		\begin{center}
      			\includegraphics[width=0.5\linewidth]{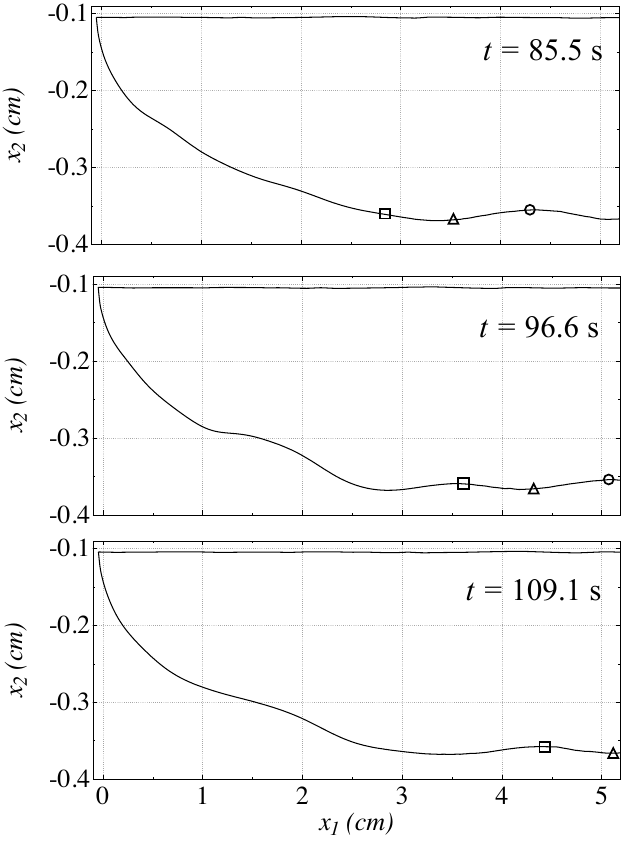}
  		\end{center}
   		\caption{Variations in sheet thickness for case 2 with $\od{\sigma}{T} = 1 \times 10^{-4} \  \mathrm{N/(m \cdot K)}$ and $u_{s,1} = 0.7\ \mathrm{mm/s}$ (See \href{https://youtu.be/B-N_8X0kYxM}{Video~2}).  }
   		\label{fig:thickness}
\end{figure}

As the solidification interface responded  to the changing flow field, the interface shape changed significantly. Deformations in the shape of the bottom of the sheet, which were often much larger than the surface corrugations, are shown in Fig.~\ref{fig:thickness}. The interface underwent large variations in shape near the TJP due to the highly unsteady flow near it. This resulted in the formation of a varying sheet thickness  as shown in Fig.~\ref{fig:thickness}. \href{https://youtu.be/B-N_8X0kYxM}{Video 2} shows a movie of the sheet thickness in a manner similar to Fig.~\ref{fig:thickness}. The variations in thickness formed near the TJP did not change significantly further downstream and were pulled with the sheet. This is shown by the markers in the figures, which translate with the pull speed and track the thickness variations. The top surface of the sheet is also shown in Fig.~\ref{fig:thickness}  where surface corrugations are barely noticeable compared to deformations on the bottom of the sheet. 

Such non-uniformities in the sheet thickness were reported by Daggolu et al.~\cite{Daggolu:2022} as a major challenge in achieving a sheet with constant thickness. To achieve their target thickness of $200 \mathrm{\mu m}$ they  added a thickness control section with several heaters after the growth section, controlled by a model-based thinning algorithm, to reduce the thickness and improve the uniformity. They carried out a few iterations to improve thickness and uniformity. Their data indicates that even after iterative improvement in the thickness control section, the standard deviation of thickness was on the same order of magnitude as the ribbon thickness. Daggolu et al.~\cite{Daggolu:2022}  did not pinpoint the main reason for thickness variations and mentioned “non-idealities in equipment, gas interaction and melt convection effects”. The numerical results show that the thickness variations are caused by the chaotic flow

\section{Conclusions}
An unsteady simulation of a horizontal ribbon growth model  including Marangoni and buoyancy effects was carried out. It was found that the combination of Marangoni and buoyancy effects causes an unsteady chaotic flow. The flow field was characterized by significant changes  driven by cold jets streaming into the crucible from the surface near the points of minimum temperature (i.e. maximum surface tension). There was often a jet just upstream of the TJP near the point with minimum supercooled temperature and one or more jets emerging from local minima in temperature further upstream. As the jets tried to follow the varying position of minimum temperature, the jets moved back and forth, interacted with each other and the rest of the flow field, and a chaotic flow field ensued.  

As the TJP position varied due to this unstable flow field, surface corrugations were formed on the top surface of the sheet. Similarly, as the interface adapted to this chaotic flow, large nonuniformities appeared on the bottom of the solid resulting in a sheet with large variations in thickness. Furthermore, the results showed sharp and large changes in growth rate at the TJP on the order of growth rate itself. These behaviors have all been observed in the experimental results of Kellerman et al.~\cite{Kellerman:2016} and Daggolu et al.~\cite{Daggolu:2022}. Thus, the chaotic flow seems to qualitatively explain most of the experimentally observed phenomena. 

Quantitatively, the median of the peak-to-peak amplitude of the surface corrugations was on the same order as the experimental values and reduced for the case corresponding to pure silicon with a greater temperature sensitivity of surface tension. Similarly, the wavelength of surface corrugations reduced with increasing  temperature sensitivity of surface tension to values on the same order as those from experiments. The dependence of amplitudes and wavelengths on pull speed was not clear. However, results suggests that only some surface wavelengths, likely due to vortical structures in the flow that had a specific turnover time, were scaled with the pull speed. Overall, given the complexity of the observed phenomena and the sensitivity of material parameters, the agreements between the experimental and model provide confidence that the observed experimental phenomena are due to Marangoni-induced flow effects.

\section{Appendix: Initial conditions and solution method}
\subsection{Initial conditions}
\label{sec:IC}
The free surface shape was initialized as:
\[ x_2 = - d_{T} e^{\frac{x_1 - x_{le}}{l_c}} \]
where the initial axial position of the triple junction was $x_{le} = -0.1d$,  $l_c = \sqrt{\frac{\sigma }{\rho g}}$ is the capillary length, and the depth of the triple junction point relative to the upper left corner of the domain (where $x_2 = 0$) was set from balance of hydrostatic pressure and surface tension as
\[ d_{T} = \sqrt{\frac{2 \sigma (1- \cos \theta_g) }{\rho g}}\]
where  $\theta_g = 11^\circ$ is the growth angle at the TJP~\cite{Surek:1976, Eustathopoulos:2010}. The solid-liquid interface shape was initialized as:
\[ x_2 = - d_{T} - t_{seed} \left( 1 - e^{-\frac{\tan(55^\circ) (x_1-x_{le})}{t_0}} \right) \]
where the initial solid sheet thickness was $t_{seed} = 0.2d$.

\subsection{Solution method}
\label{sec:SM}
An initial steady solution was obtained by fixing the solid-liquid interface, $\od{\rho}{T} = \od{\sigma}{T}=0$, and using linear basis functions. Then, an adaptive time-stepping was used to obtain a steady solution while the ALE moving mesh method and mesh adaptation tracked the interface and kept the mesh quality and density. Next, a steady solution was obtained using quadratic and then quartic basis functions (p-refinement). Then, the mesh adaptation refined the solution based on a target error (h-refinement). Marangoni stress was next gradually increased up to  $\od{\sigma}{T} = 1 \times 10^{-4} \  \mathrm{N/(m \cdot K)}$. Except for case 1 a steady solution was obtained at $\od{\sigma}{T} = 1 \times 10^{-4} \  \mathrm{N/(m \cdot K)}$. Next, the temperature sensitivity of surface tension was set to  $1 \times 10^{-4} \  \mathrm{N/(m \cdot K)}$ and $\od{\rho_l}{T} = -0.23\ \mathrm{kg/(m^3 \cdot K)}$ and  an adaptive time-stepping was used for a period of $\frac{4d}{u_{s,1}}$ to let the effects of imposed buoyancy and Marangoni in the flow  be established. Finally, a maximum time step of $\Delta t = \frac{l_{min}}{u_{s,1}}$ was set. The adaptive time-stepping used in this stage reduced the time step by factors of two, if needed for convergence, such that results were always obtained at $\Delta t$ intervals. 

 \section*{Acknowledgements}
This material is based upon work supported by the National Science Foundation under Grant No. 1762802.

\bibliographystyle{ieeetr}
\bibliography{_dendrite.bib}

\end{document}